\documentclass[preprint,review, 12pt]{elsarticle}

\usepackage{amssymb}
\usepackage{subfigure}

\usepackage{lineno}
\usepackage{color}
\usepackage{gensymb}
\usepackage[colorlinks=true, allcolors=blue]{hyperref}

\journal{NIMA}

\newcommand{\Lcms}{\mathrm{cm^{-2}s^{-1}}}
\newcommand{\Neqcm}{\mathrm{n_{eq}cm^{-2}}}
\newcommand{\Ifb}{\mathrm{fb^{-1}}}
\newcommand{\Vneq}{\mathrm{V_{n_{eq}}}}

\begin{document}

\begin{frontmatter}

\title{Radiation Campaign of HPK Prototype LGAD sensors for the High-Granularity Timing Detector (HGTD)
}

\author[label1]{X. Shi\corref{cor}}
\ead{shixin@ihep.ac.cn}
\cortext[cor]{Corresponding author}


\author[label1]{M. K. Ayoub}
\author[label1]{J. G. da Costa}
\author[label1,label2]{H. Cui}
\author[label1]{R. Kiuchi}
\author[label1]{Y. Fan}
\author[label1,label2]{S. Han}
\author[label1]{Y. Huang}
\author[label1,label2]{M. Jing}
\author[label1]{Z. Liang}
\author[label1]{B. Liu}
\author[label1]{J. Liu}
\author[label1]{F. Lyu}
\author[label1,label2]{B. Qi}
\author[label1,label2]{K. Ran}
\author[label1]{L. Shan}
\author[label1]{L. Shi}
\author[label1,label2]{Y. Tan}
\author[label1,label2]{K. Wu}
\author[label1,label2]{S. Xiao}
\author[label1,label2]{T. Yang}
\author[label1]{Y. Yang}
\author[label1,label2]{C. Yu}
\author[label1]{M. Zhao}
\author[label1]{X. Zhuang}

\address[label1]{State Key Laboratory of Particle Detection and Electronics, Institute of High Energy Physics, Chinese Academy of Sciences, 19B Yuquan Road, Shijingshan District, Beijing 100049, China}
\address[label2]{University of Chinese Academy of Sciences, 19A Yuquan Road, Shijingshan District, Beijing 100049, China}

\author[label11]{L. Castillo Garc\'{i}a}
\author[label11]{E. L. Gkougkousis}
\author[label11]{C. Grieco}
\author[label11]{S. Grinstein}

\address[label11]{Institut de F\'{i}sica d'Altes Energies (IFAE), Carrer Can Magrans s/n, Edifici Cn, Universitat Aut\'{o}noma de Barcelona (UAB), E-08193 Bellaterra (Barcelona), Spain}


\author[label14]{M. Leite}
\author[label14]{G. T. Saito}

\address[label14]{Instituto de F\'{i}sica - Universidade de S\~{a}o Paulo (USP), R. do Mat\~{a}o, 1371, Cidade Universit\'{a}ria, S\~{a}o Paulo - SP 05508-090 - Brazil}


\author[label15]{A. Howard}
\author[label15]{V. Cindro}
\author[label15]{G. Kramberger}
\author[label15]{I. Mandi\'{c}}
\author[label15]{M. Miku$\check{z}$}

\address[label15]{Jozef Stefan Institut (JSI), Dept. F9, Jamova 39, SI-1000 Ljubljana, Slovenia}

\author[label9]{G. d'Amen}
\author[label9]{G. Giacomini}
\author[label9]{E. Rossi}
\author[label9]{A. Tricoli}

\address[label9]{Brookhaven National Laboratory (BNL), Upton, NY 11973, U.S.A.}


\author[label7]{H. Chen}
\author[label7]{J. Ge}
\author[label7]{C. Li}
\author[label7]{H. Liang}
\author[label7]{X. Yang}
\author[label7]{L. Zhao}
\author[label7]{Z. Zhao}
\author[label7]{X. Zheng}

\address[label7]{Department of Modern Physics and State Key Laboratory of Particle Detection and Electronics, University of Science and Technology of China, Hefei 230026, China}


\author[label5]{N. Atanov}
\author[label5]{Y. Davydov}

\address[label5]{Joint Institute for Nuclear Research, Joliot-Curie street 6, Dubna, 141980 Russia}


\author[label12]{J. Grosse-Knetter}
\author[label12]{J. Lange}
\author[label12]{A. Quadt}
\author[label12]{M. Schwickardi}

\address[label12]{II. Physikalisches Institut, Georg-August-Universit\"{a}t, Friedrich-Hund-Platz 1, 37077 G\"{o}ttingen, Germany}


\author[label3]{S. Alderweireldt}
\author[label3]{A. S. C. Ferreira}
\author[label3]{S. Guindon}
\author[label3]{E. Kuwertz}
\author[label3]{C. Rizzi}

\address[label3]{CERN, Esplanade des Particules 1, 1211 Geneva 23 }


\author[label8]{S. Christie}
\author[label8]{Z. Galloway}
\author[label8]{C. Gee}
\author[label8]{Y. Jin}
\author[label8]{C. Labitan}
\author[label8]{M. Lockerby}
\author[label8]{S. M. Mazza}
\author[label8]{F. Martinez-Mckinney}
\author[label8]{R. Padilla}
\author[label8]{H. Ren}
\author[label8]{H. F.-W. Sadrozinski}
\author[label8]{B. Schumm}
\author[label8]{A. Seiden}
\author[label8]{M. Wilder}
\author[label8]{W. Wyatt}
\author[label8]{Y. Zhao}

\address[label8]{SCIPP, Univ. of California Santa Cruz, CA 95064, USA}


\author[label13]{D. Han}
\author[label13]{X. Zhang}

\address[label13]{Novel Device Laboratory, Beijing Normal University, No. 19, Xinjiekouwai Street, Haidian District, Beiing 100875, China}

\begin{abstract}
    We report on the results of a radiation campaign with neutrons and protons 
    of Low Gain Avalanche Detectors (LGAD) produced by Hamamatsu (HPK) as prototypes for the High-Granularity Timing Detector (HGTD) in ATLAS. 
    Sensors with an active thickness of 50~$\mu$m were irradiated in steps of roughly 2$\times$ up to a fluence of $3\times10^{15}~\Neqcm$. As a function of the fluence, the collected charge and time resolution of the irradiated sensors will be reported for operation at -30\celsius.


\end{abstract}



\begin{keyword}
Low Gain Avalanche Detector \sep Radiation \sep Timing detector 


\end{keyword}

\end{frontmatter}


\section{Introduction}

The High-Luminosity Large Hadron Collider (HL-LHC) will reach the instantaneous luminosity up to $7.5 \times 10^{34}~\Lcms$ around 2027. The ATLAS experiment is scheduled to upgrade during the Long Shutdown 3 (LS3) from 2025 to 2027 to cope with the higher luminosities. One direct consequence of the high luminosity is the multiple collisions for single bunch crossing, so-called pile-up events. To mitigate the pile-up effects, ATLAS proposed the design of a High Granularity Timing Detector (HGTD) to enhance the Phase-II silicon-based Inner Tracker (ITk)~\cite{2017-ATLTDR-ITkPix, 2017-ATLTDR-ITkStp} in the forward region. The target average time resolution for a minimum-ionising particle (MIP) is 30~ps to 50~ps per track during the HL-LHC operation. 

\begin{figure}[hptb]
    \centering{
        \includegraphics[width=0.8\textwidth]{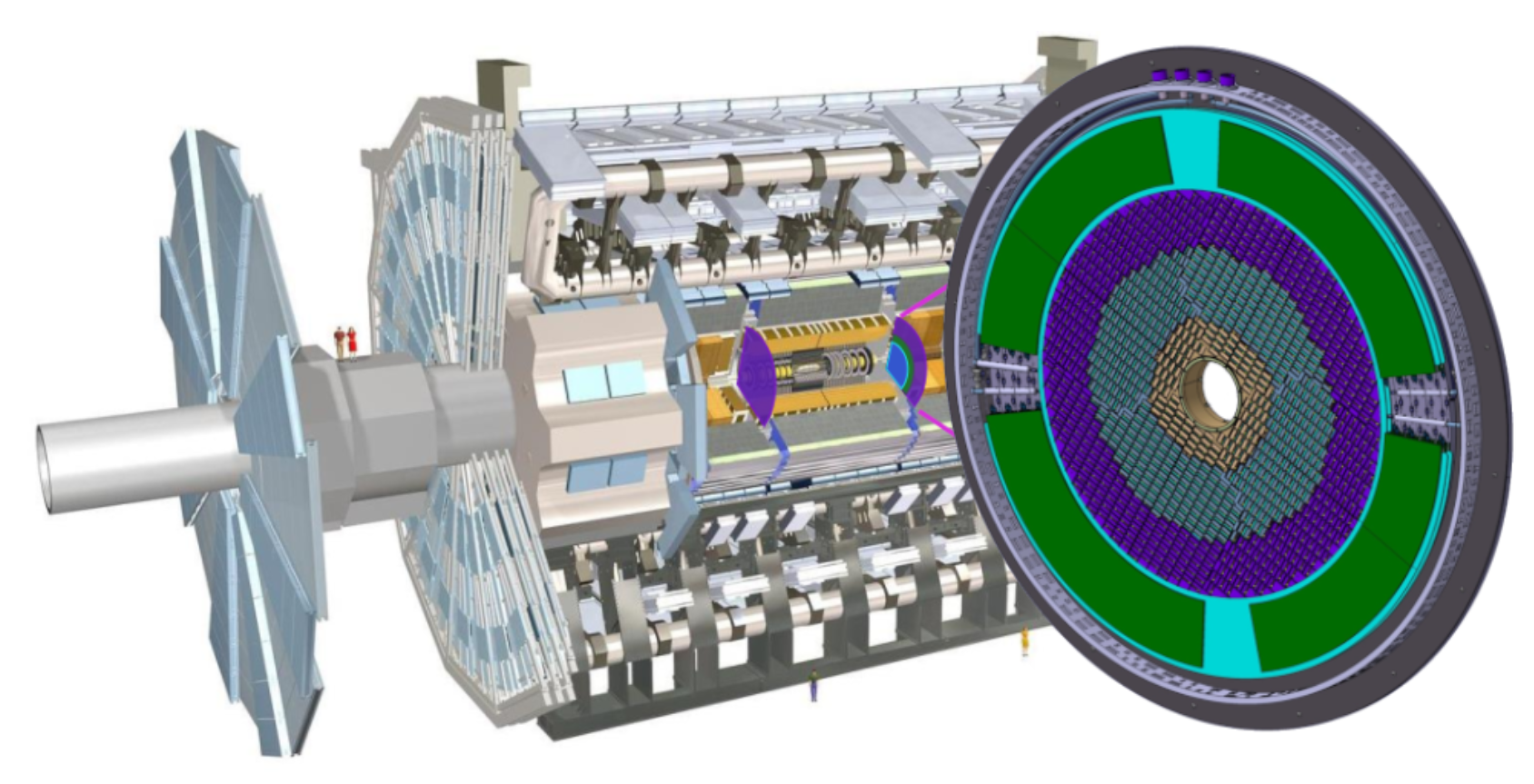}
        \caption{Position of the HGTD within the ATLAS detector. The HGTD acceptance is defined as the surface covered by the HGTD between a radius of 12~cm and 64~cm (the three inner rings) at a position of z = $\pm$3.5~m along the beamline, on both sides of the detector. The green (color online) ring are the peripheral electronics.} 
        \label{fig:atlas-hgtd}
    }
\end{figure}

The nominal HL-LHC operation will have around 200 pile-up for the same bunch crossing. The HGTD is designed to cope with 200 pile-up for a total integrated luminosity of 4000~$\Ifb$. It will be located in the gap between the barrel and the end-cap calorimeters, at $\pm$3.5~m from the interaction point. The two vessels outside the ITk volume are shown in Fig.~\ref{fig:atlas-hgtd}. 


As the HGTD is placed in the forward region of ATLAS detector, it is crucial for the sensors and electronics to maintain adequate radiation hardness. For the data taking of the HL-LHC running, the maximum nominal neutron-equivalent fluence at a radius of 12~cm should reach $5.6\times10^{15}~\Neqcm$ and the total ionizing dose (TID) about 3.3~MGy. With a safety factor of 1.5 for the sensors and 2.25 for the electronics which are more sensitive to the TID, the detector would need to withstand $8.3\times10^{15}~\Neqcm$ and 7.5~MGy. 

\begin{figure}[hptb]
    \centering{
        \includegraphics[width=0.45\textwidth]{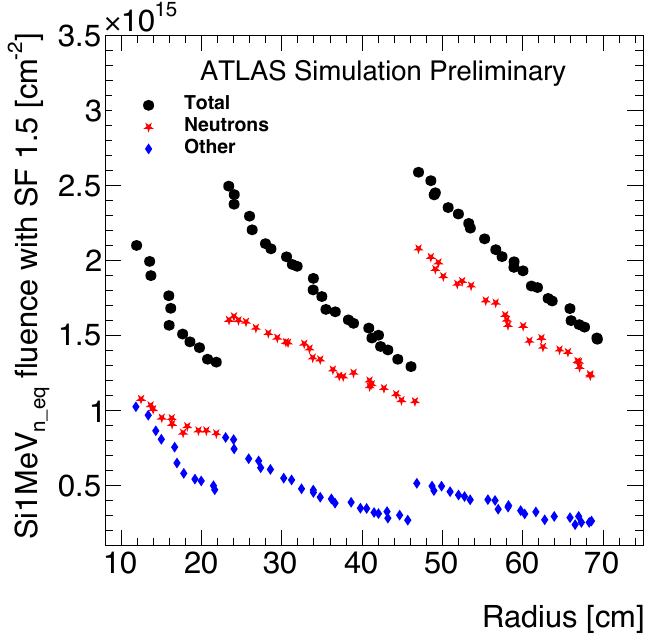}
        \includegraphics[width=0.45\textwidth]{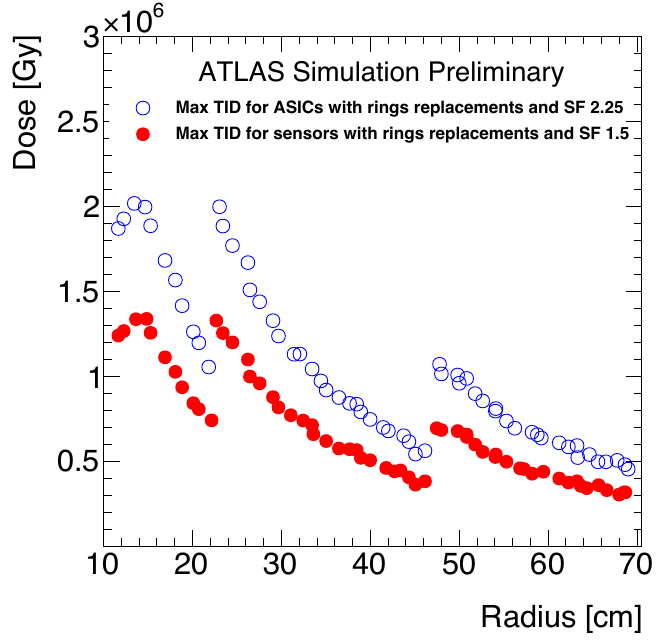}
        \caption{
        Simulated neutron-equivalent fluence (left) and total ionizing dose (right). Expected Silicon (Si) 1~Me$\Vneq$ radiation levels in HGTD, using Fluka simulations, as a function of the radius considering a replacement of the inner ring every 1000~$\Ifb$ and the middle ring replaced at 2000~$\Ifb$. For the radiation levels, the particle type is included and the contribution from charged hadrons is included in `Others'. These curves include a factor of 1.5 to account for simulation uncertainty. An additional factor of 1.5 is applied to the TID to account for low dose rate effects on the electronics, leading to a SF = 2.25.}
        \label{fig:rad-level}
    }
\end{figure}

The HGTD is designed with 3 rings as ``inner", ``middle", and ``outer" with boundaries from 12-23~cm, 23-47~cm, and 47-64~cm, respectively. The inner ring is designed to be replaced every 1000~$\Ifb$, the middle ring to be replaced at 2000~$\Ifb$, and the outer ring never to be replaced until 4000~$\Ifb$. A simulation of the radiation levels for three rings is shown in Fig.~\ref{fig:rad-level}. It can be seen that the maximum fluence and TID will be $2.5\times10^{15}~\Neqcm$ and 2.0~MGy respectively. The final radial transition between the three rings for the complete detector layout will be done after the ITk layout and the re-evaluation of radiation hardness of the final sensors and ASICs.

\section{Low Gain Avalanche Detectors (LGAD)}

To meet the above mentioned irradiation requirement, the selected technology for the HGTD sensors is Silicon Low Gain Avalanche Detectors (LGAD) with a baseline active thickness of 50~$\mu$m. The target gain (charge) for a MIP is 20 (10~fC) at the start and 8 (4~fC) at the end of lifetime. 

\begin{figure}[hptb]
    \centering{
        \includegraphics[width=0.6\textwidth]{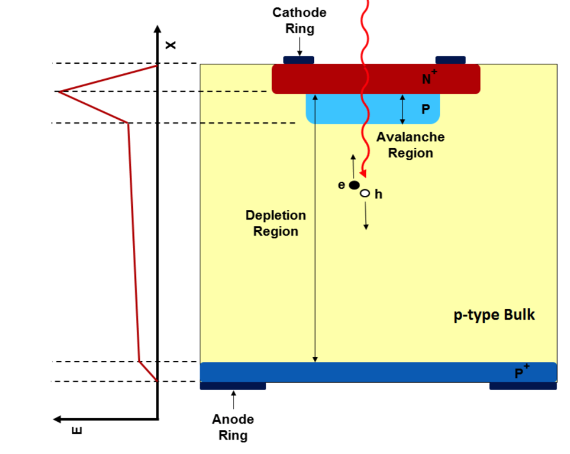}
        \caption{Sketch of the cross section of a LGAD with a charged particle passing through the gain layer. On the left side a qualitative profile of the electrical field amplitude is shown, where the peak is located in the same region of gain layer in which the avalanche happens.}
        \label{fig:lgad-schem}
    }
\end{figure}

LGADs are segmented planar Silicon detectors with internal gain~\cite{2018-RPP-4D-Tracking-UFSD} as sketched in Fig.~\ref{fig:lgad-schem}. 
The design is based on a modification of the doping profile where an additional doping layer of $p^+$ material (Boron or Gallium) is introduced close to the n-p junction, which will give a charge gain of 10-20~\cite{2014-NIMA-Tech-LGAD}. Since the multiplication mechanism starts for electrons at a lower value of the electric field than holes and since p-bulk material is more radiation hard, the n-in-p design offers the best control and better stability over the multiplication process. 

The LGADs are pioneered by the Centro Nacional de Microelectr\'{o}nica (CNM, Barcelona)~\cite{2014-NIMA-Tech-LGAD} and developed within CERN-RD50 community~\cite{RD50} also in collaboration with LGAD vendors as Hamamatsu Photonics (HPK, Japan) and Fondazione Bruno Kessler (FBK, Italy).  Other sites such as Micron (UK), Brookhaven National Lab (BNL, US), and Novel Device Laboratory (NDL, China) are also in the process of producing the sensors.

In this paper, we concentrate on one HPK prototype LGAD production, labeled HPK-3.2. It has an active thickness of 50 $\mu$m and depth of 2.2 $\mu$m gain layer with Boron dopant. To avoid duplication, more details about its layout, doping and pre-rad performance can be found in~\cite{2020-CDS-HPK-Layout-Perform}.  


\section{Radiation damage and irradiations}

In the innermost region of HGTD (r~\textless~23~cm), the radiation field is roughly equal for neutrons and charged hadrons, while in the outer regions the field is dominated by neutrons due to backscatter from the adjacent calorimeters. The maximum fluence from charged particles is around $1\times10^{15}~\Neqcm$ and for neutron $2\times10^{15}~\Neqcm$. 

Radiation damage in Silicon mainly results in the change of the effective doping concentration, increase of the leakage current and charge loss due to trapping and recombination~\cite{2005-NIMA-Rad-Hard-SLHC}. For LGADs, one of the main effects is the degradation of gain 
due to acceptor removal effect~\cite{2015-JINST-Rad-LGAD, 2018-NIMA-Thin-LGAD}. 

To study the LGAD performance after irradiation, sensors have been irradiated up to a fluence of $3\times10^{15}~\Neqcm$ at various facilities with different particle types and energies. The facilities relevant in this paper are JSI~\cite{2006-NIMA-TRIGA} in Ljubjana with 1 MeV neutron (hardness factor 0.9, uncertainty 10\%) and CYRIC~\cite{2015-JINST-CYRIC} in Japan with 70 MeV proton (hardness factor 1.5).

\section{LGAD performance before and after irradiation}

The HPK LGAD sensors have been tested before and after irradiation by various HGTD groups. The pre-irradiation measurements including current-voltage (I-V) and capacitance-voltage (C-V) characteristics as well as the distance between the two multiplication layers of adjacent pads (inter-pad gap) regions can be found in~\cite{2020-CDS-HPK-Layout-Perform}. 

The charge collection of the LGADs are measured with $^{90}\mathrm{Sr} ~\beta$-source (setup see Sec.~4 in Ref.~\cite{2019-NIMA-HPK-UFSD}) a read-out board with a 4700 $\Omega$ trans-impedance developed by UCSC~\cite{2017-NIMA-Beam-16ps-UFSD}. The laboratory setup consists of alignment frame and climatic chamber than can be operated at $-30\celsius$. The trigger and time reference is provided by another well-calibrated LGAD with a resolution of 35 ps at room temperature~\cite{2019-NIMA-HPK-UFSD}.

\begin{figure}[hptb]
    \centering{
        \includegraphics[width=0.75\textwidth]{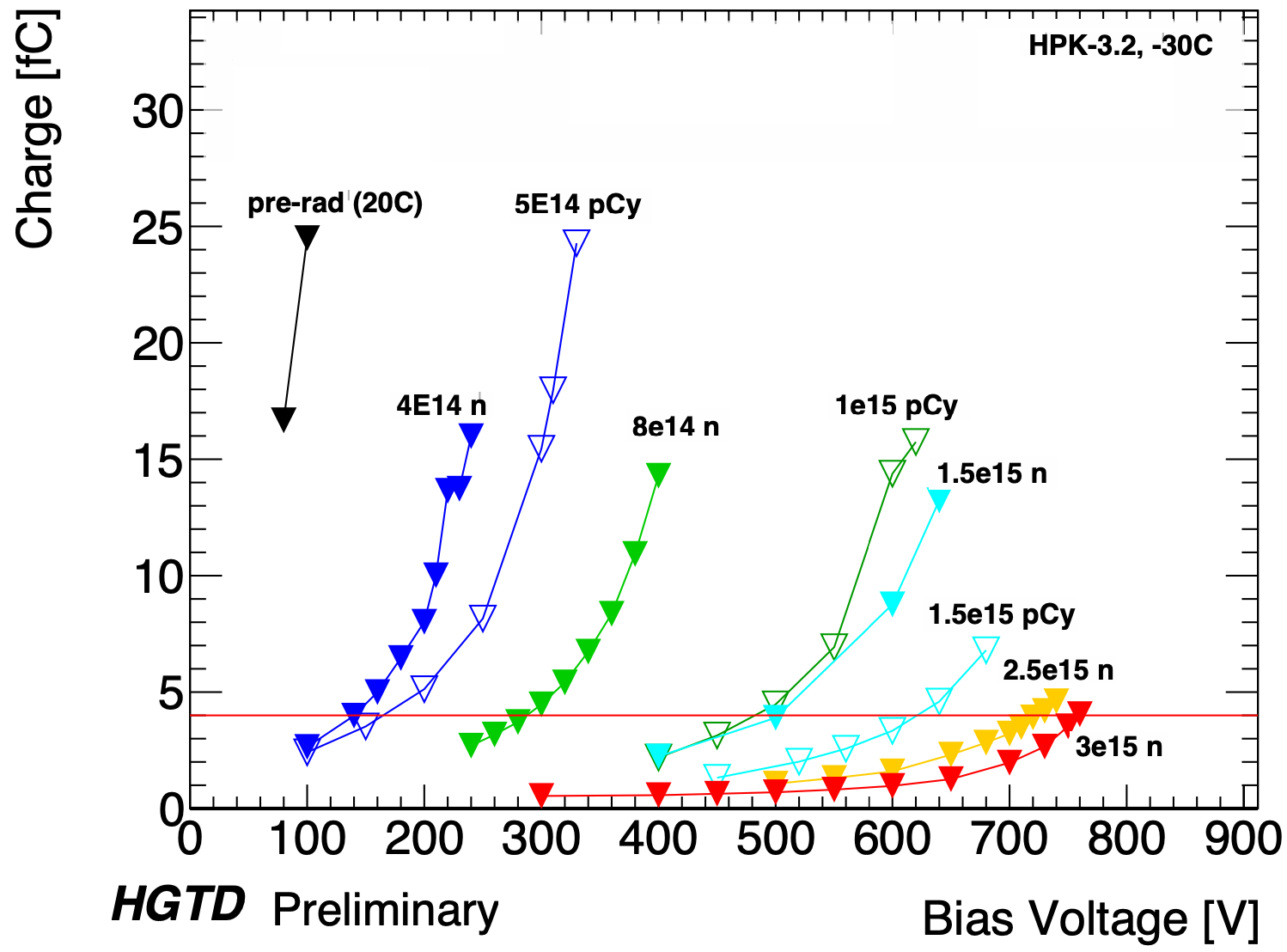}
        \caption{Collected charge as a function of bias voltage for different fluences for HPK-3.2. The horizontal line indicates the HGTD lower charge limit of 4~fC at all fluences. Solid markers indicate n irradiation (\textit{n}), open markers p irradiation at CYRIC (\textit{pCy}). The numbers in the legend are the fluences with unit $\Neqcm$, and C stands for \celsius. Measurements were performed at -30\celsius ~except for the pre-rad measurement that was done at 20\celsius. } 
        \label{fig:charge-bias}
    }
\end{figure}

\begin{figure}[hptb]
    \centering{
        \includegraphics[width=0.75\textwidth]{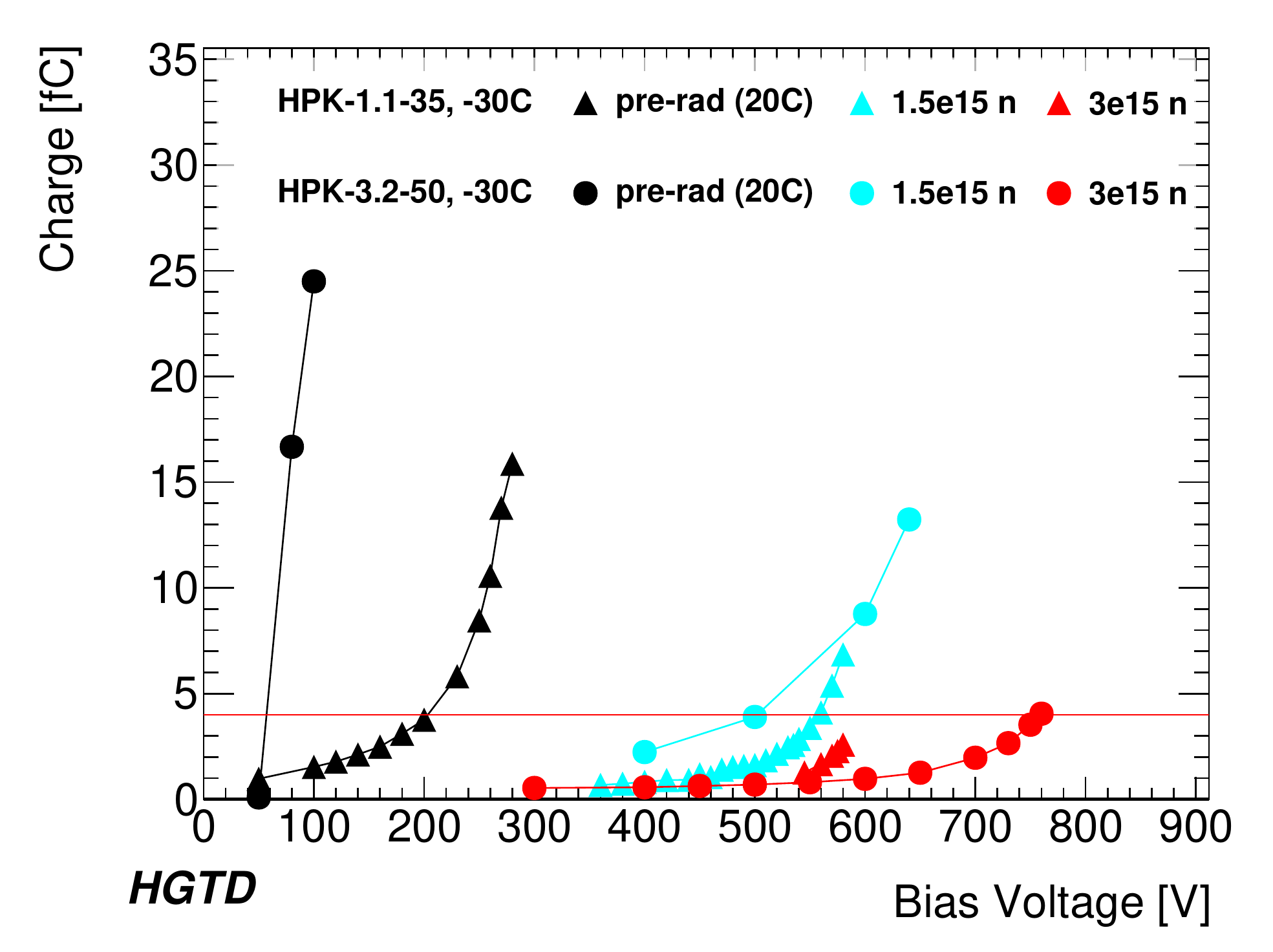}
        \caption{Collected charge as a function of bias voltage for different fluences for HPK-1.1 (active thickness 35~$\mu$m) and HPK-3.2 (active thickness 50~$\mu$m). The horizontal line indicates the HGTD lower charge limit of 4~fC at all fluences. The HPK-1.1 did not satisfy the charge requirement for highest fluence. Solid markers indicate n irradiation (\textit{n}). The numbers in the legend are the fluences with unit $\Neqcm$, and C stands for \celsius. Measurements were performed at -30\celsius ~except for the pre-rad measurement that was done at 20\celsius.  
        } 
        \label{fig:hpk11_hpk32}
    }
\end{figure}

Fig.~\ref{fig:charge-bias} shows the collected charge as a function of bias voltage before and after neutron and proton irradiation up to $3\times10^{15}~\Neqcm$ for HPK-3.2. It is evident that before irradiation, a much lower bias is sufficient to collect a higher charge. As the fluence level goes higher, the sensor's capability to collect charge is dropped even with higher bias voltage because of the reduction of space charge density in the gain layer due to the acceptor removal effect. A stronger acceptor removal effect for proton irradiation than neutron irradiation for the same fluence is also noticeable due to the higher fraction of point defects in the proton irradiation~\cite{2018-IEEE-DISP-DAMG}. A charge of 4~fC was found to be the lower limit that still satisfies the HGTD science requirements. For comparison, a previous prototype HPK-1.1 with 35~$\mu$m active thickness did not satisfy the 4~fC collected charge requirement at the highest fluence as shown in Fig.~\ref{fig:hpk11_hpk32}. 

It had been observed that sensors irradiated with proton show higher acceptor removal rate than neutron irradiated sensors at the same fluence \cite{2015-JINST-Rad-LGAD, 2018-NIMA-Thin-LGAD}, which is also consistent the results shown in Fig.~\ref{fig:charge-bias}. This studies will be followed up by irradiations with higher energy charged hadrons and mixed neutron-proton irradiations for a realistic estimation with the expected final particle composition. 


The time resolution $\sigma_t$ is determined by: $\sigma_t^2 = \sigma^2_{Landau ~noise} + \sigma^2_{Jitter} + \sigma^2_{Signal~distortion} + \sigma^2_{Time~walk}$, where the main contributions are the two leading terms. The ``Landau noise'' is caused by the non-uniform energy deposition of ionization. The jitter is proportional to the rise time, the noise and the inverses of the gain. The signal distortion and the time walk can be mitigated by the geometry of sensor and the constant fraction discriminator (CFD) method respectively. For detailed study of these terms can be found in \cite{2018-RPP-4D-Tracking-UFSD} and various beam tests~\cite{2017-NIMA-Beam-16ps-UFSD, 2017-JINST-LGAD, 2018-JINST-Beam-LGAD}. 

\begin{figure}[hptb]
    \centering{
        \includegraphics[width=0.8\textwidth]{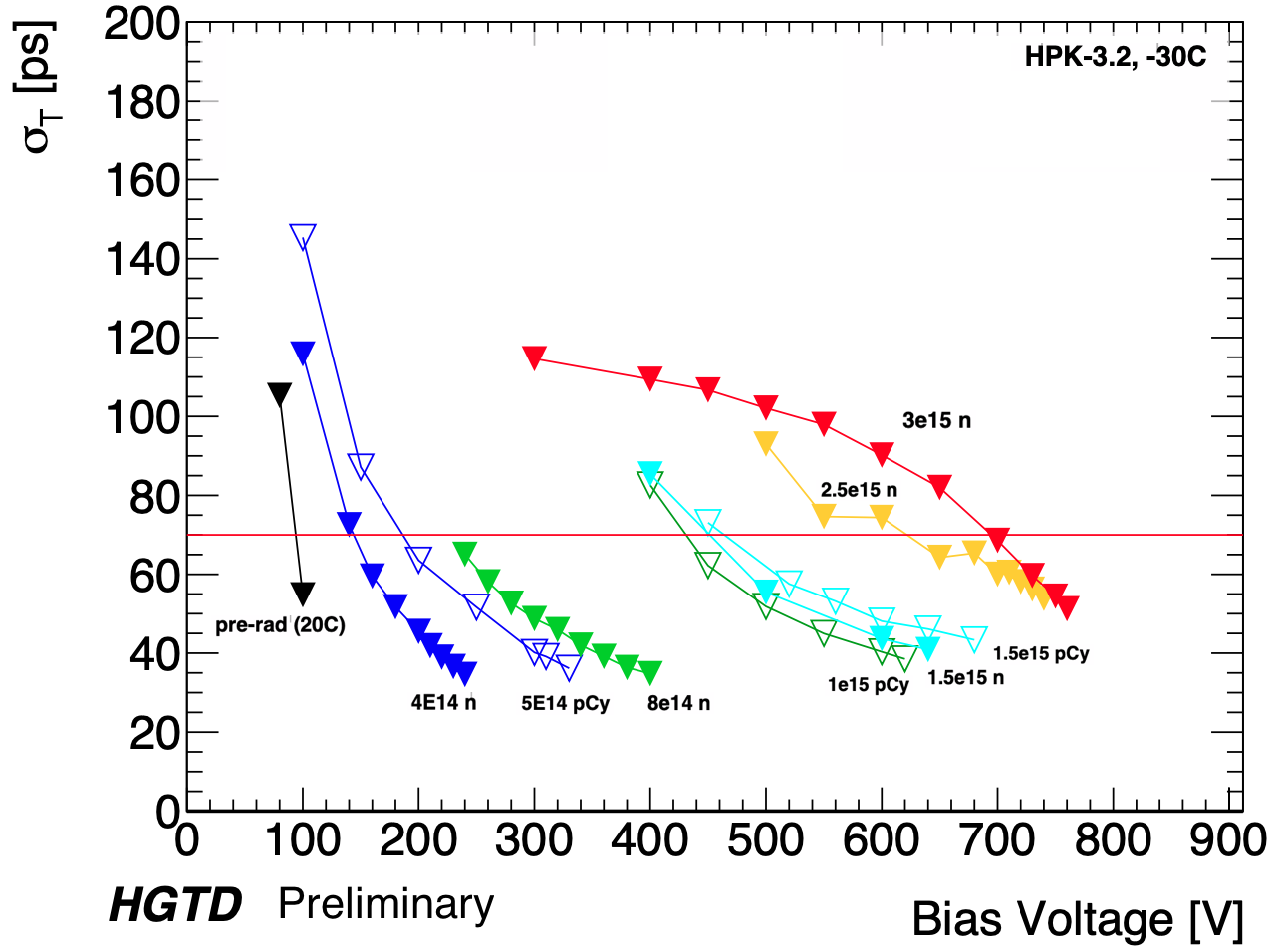}
        \caption{Time resolution as a function of bias voltage for different fluences for HPK-3.2 sensors measured on custom-made HGTD-specific readout boards. Solid markers indicate n irradiation (\textit{n}), open markers p irradiation at CYRIC (\textit{pCy}). The red line represents the maximum allowed time resolution (70~ps) in the lifetime of HGTD. The numbers in the legend are the fluences with unit $\Neqcm$, and C stands for \celsius. Measurements were performed at -30\celsius ~except for the pre-rad measurement that was done at 20\celsius.} 
        \label{fig:time}
    }
\end{figure}

Time resolution of HPK-3.2 was measured in the $\beta$-telescope after irradiation with 1 MeV neutrons at JSI and 70 MeV protons at CYRIC. The results shown in Fig.~\ref{fig:time} indicates that a time resolution of less than 70~ps can be achieved up to the highest fluence of $2.5\times10^{15}~\Neqcm$ to fulfill the requirement of HGTD but with bias voltages larger than 700V for the highest fluences. The reduced time resolution performance for HPK-3.2 before irradiation was mainly due to early breakdown and the sensor has to be operated at low bias which resulted in non-saturated carrier velocity~\cite{2020-CDS-HPK-Layout-Perform}.  

 It is also noticeable the curvature of the time resolution drop for $2.5\times10^{15}~\Neqcm$ and $3\times10^{15}~\Neqcm$ neutron are different than others. The exact reason is unknown, but with more irradiation fluence in the future, it will be an interesting point to follow up. 

\section{Summary}

In order to cope with the pile-up challenge of HL-LHC, the LGAD sensors are proposed to be used for the ATLAS HGTD, with radiation tolerance requirement of $2.5\times10^{15}~\Neqcm$ of fluence and 2 MGy of TID. One LGAD prototype developed by HPK with thickness 50~$\mu$m has been irradiated up to the $3\times10^{15}~\Neqcm$. The measured charge collection of 4~fC and time resolution of better than 70~ps at the highest fluence fulfill the operation requirement of HGTD. Further studies will be performed with LGADs from other vendors as well as dedicated readout ASICs.

\section*{Acknowledgement}
This work was supported by the United States Department of Energy,
grant DE-FG02-04ER41286, “the Fundamental Research Funds for the Central Universities” of China (grant WK2030040100), the National Natural Science Foundation of China (No. 11961141014), the State Key Laboratory of Particle Detection and Electronics (SKLPDE-ZZ-202001), the Hundred Talent Program of the Chinese Academy of Sciences (Y6291150K2), the CAS Center for Excellence in Particle Physics (CCEPP), the MINECO, Spanish Government, under grant RTI2018-094906-B-C21, the Slovenian Research
Agency (project J1-1699 and program P1-0135), the U.S.A. Department of Energy under grant contact DE-SC0012704 and partially carried out at the USTC Center for Micro and Nanoscale Research and Fabrication and partially performed within the CERN RD50 collaboration. The contributions from UCSC technical staff and students is acknowledged.

\bibliographystyle{unsrt}
\bibliography{p7_lgad_hpk}





\end{document}